# Klein Tunneling of Gigahertz Elastic Waves in Nanoelectromechanical Metamaterials

Daehun Lee[1,†], Yue Jiang[2,†], Xiaoru Zhang[1], Shahin Jahanbani[1,3], Chengyu Wen[4], Qicheng Zhang[5,6***], A. T. Charlie Johnson[2,7**], Keji Lai[1,*]

[1] Department of Physics, University of Texas at Austin, Austin, TX, USA

[2] Department of Physics and Astronomy, University of Pennsylvania, Philadelphia, PA, USA

[3] Present address: Department of Physics, University of California, Berkeley, CA, USA

[4] Department of Electrical and Systems Engineering, University of Pennsylvania, Philadelphia, PA, USA

[5] Research Center for Industries of the Future, Westlake University, Hangzhou, Zhejiang 310030, China

[6] Key Laboratory of 3D Micro/Nano Fabrication and Characterization of Zhejiang Province, School of Engineering, Westlake University, Hangzhou, Zhejiang 310030, China

[7] Department of Materials Science and Engineering, University of Pennsylvania, Philadelphia, PA, USA

[†] These authors contributed equally to this work

* E-mails: zhangqicheng@westlake.edu.cn; cjohnson@physics.upenn.edu; kejilai@physics.utexas.edu

## Summary

Klein tunneling, the perfect transmission of a normally incident relativistic particle through an energy barrier, has been tested in various electronic, photonic, and phononic systems. Its potential in guiding and filtering classical waves in the Ultra High Frequency regime, on the other hand, has not been explored. Here, we report the realization of acoustic Klein tunneling in a nanoelectromechanical metamaterial system operating at gigahertz frequencies. The piezoelectric potential profiles are obtained by transmission-mode microwave impedance microscopy, from which reciprocal-space maps can be extracted. The transmission rate of normally incident elastic waves is near unity in the Klein tunneling regime and drops significantly outside this frequency range, consistent with microwave network analysis. Strong angular dependent transmission is also observed by controlling the launching angle of the emitter interdigital transducer. This work broadens the horizon in exploiting high-energy-physics phenomena for practical circuit applications in both classical and quantum regimes.

# Introduction

Because of their underlying wave nature, electromagnetic waves in photonic systems and acoustic / elastic waves in phononic systems share many common features with electronic waves in condensed matter systems. As a result, the frequency dispersion of engineered metamaterials closely resembles the band structure of solid-state materials, which has been extensively studied in recent years[1,2]. For instance, macroscopic periodic structures can be constructed to test exotic phenomena such as Majorana zero modes[3,4] and non-Hermitian bands[5,6], which are difficult to achieve in real materials. Conversely, quantum physics in condensed matter systems can provide deep insight on the realization of unusual functionalities in various metamaterials, such as topologically protected transport against structural defects or sharp bends[7-10]. In this context, the classical analogue to the Klein tunneling effect[11-13] – unity transmission of a relativistic particle passing through a potential barrier upon normal incidence – holds high promise for waveguiding and filtering applications in photonic and phononic systems.

In elementary quantum mechanics, the transmission probability of a nonrelativistic particle decays exponentially when passing through a classically forbidden region[14]. For relativistic Dirac particles, however, normal transmission is unimpeded regardless of the height and width of the energy barrier, an effect known as the Klein tunneling[11-13]. This counterintuitive process relies on two conditions: (1) the presence of a continuum of negative-kinetic-energy states inside the barrier that matches energy of the incoming states and (2) the conservation of chiral pseudospin $1/2$ in a bi-spinor wavefunction, which prohibits backscattering under normal incidence. Because of the difficulty in reaching relativistic limit in a particle-physics setting, experimental investigations of Klein tunneling have been mostly carried out in graphene[15,16], where quasiparticles near the charge-neutrality point behave as two-dimensional massless Dirac fermions[17,18]. Signatures of Klein tunneling in this condensed matter platform include the excess resistance across a ballistic P-N junction due to the collimation effect[19-21] and the half-period shift in magneto-conductance oscillations induced by quantum interference between two parallel interfaces[22]. Similar theoretical and experimental works have also been performed in photonic systems at optical frequencies[23,24] and phononic systems at the kilohertz (kHz) range[25-28]. It is widely accepted that analogues of the Klein tunneling effect, originally proposed as a high-energy physics paradox, can be utilized for enhancing quantum transport and manipulating light and sound waves.

In this work, we further extend the experimental investigation of phononic Klein tunneling to the gigahertz (GHz) regime, which is technologically important for wireless communication and integrated circuit applications. The nanoelectromechanical phononic metamaterials are fabricated on freestanding AlN membranes, where graphene-like structures are patterned. Using transmission-mode microwave impedance microscopy (TMIM)[29,30], we have visualized the real-space profile of the elastic waves, from which phononic band structures in the reciprocal space can be obtained by a Fourier transform. Upon normal incidence, near unity transmission is observed at the center of the Klein tunneling regime, while wave propagation outside this frequency range is strongly attenuated. By controlling the launching angle from interdigital transducers (IDTs), we also observe strong angular dependent transmission through the heterostructure, in agreement with theoretical predictions. Our work represents the first successful demonstration of Klein tunneling in integrated phononic circuits in the Ultra High Frequency (UHF, 300 MHz – 3 GHz) regime, which is desirable for classical signal processing and quantum information systems.

## Results and Discussions

The phononic crystal (PnC) structure implemented in this work is based on an acoustic analogue of graphene. As illustrated in Fig. 1a, the metamaterial is formed by etching away snowflake-like patterns[31] from suspended polycrystalline AlN films whose normal direction is along the c-axis defined in conventional crystallography. The choice of AlN is mainly due to its isotropic properties for implementing acoustic graphene. The freestanding film can minimize acoustic leakage to the substrates. Details of the sample quality and the effect of imperfections on the PnC band structure are included in Supplemental Information Section S1. The structure has the $C_{6v}$ symmetry (sixfold rotations about the center and mirror symmetry about the vertical planes) that is essential for low-energy Dirac dispersion in the band structure. Similar to our earlier work[32], we employ TMIM to study wave propagation on AlN membranes through the piezoelectric effect. As shown in Fig. 1b, elastic waves at around 1 GHz are launched by the emitter IDT, which is tilted from the normal of the PnC by an angle $\alpha$ such that elastic waves inside the PnC can propagate in the horizontal direction. Details of the momentum matching condition that determines the launching angle can be found in Supplemental Information Section S2. The piezoelectric surface potential on AlN is detected by the cantilever probe, amplified by the TMIM electronics, and demodulated by an in-phase/quadrature (I/Q) mixer[29,30]. Note that the technique is detecting the net effect of piezoelectric transduction from the vector (both out-of-plane and in-plane) displacement fields to the GHz electrical potential on the sample surface. Signals at the RF and LO ports of the I/Q mixer can be expressed as $V_{\text{RF}} \propto e^{i(\omega t - \vec{k} \cdot \vec{r})}$ and $V_{\text{LO}} \propto e^{i\omega t}$, respectively, where $\omega$ is the angular frequency and $\vec{k}$ is the wave vector. The two outputs channel of the I/Q mixer are therefore $V_{\text{Ch1}} \propto \text{Re}(V_{\text{RF}} V_{\text{LO}}^*) = \cos(\vec{k} \cdot \vec{r})$ and $V_{\text{Ch2}} \propto \text{Im}(V_{\text{RF}} V_{\text{LO}}^*) = -\sin(\vec{k} \cdot \vec{r})$. Fig. 1c and 1d show the simultaneously taken atomic-force microscopy (AFM) and TMIM-Ch1 images, respectively (complete data in Supplemental Section S3). It is clear that the TMIM is measuring the GHz potential rather than the surface topography. More importantly, by combining the two TMIM output channels as $V_{\text{Ch1}} + i * V_{\text{Ch2}}$, we obtain phase-sensitive signals that are proportional to the displacement fields, from which the reciprocal-space ($k$-space) information can be extracted through Fast-Fourier Transformation (FFT). Such an approach will be extensively exploited in our analysis below.

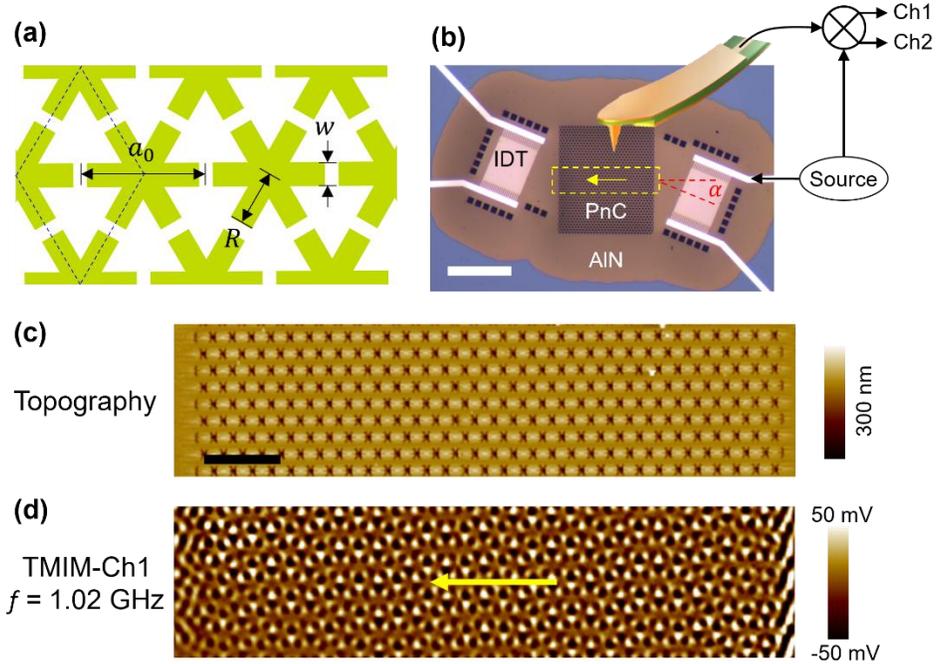

**Fig. 1| Phononic crystal design and TMIM experiment**. **a**, Schematic of the phononic crystal design, where the green snowflake regions are etched away. The dashed rhombus depicts the unit cell. **b**, Optical image of the freestanding AlN device overlaid with the TMIM setup. The normal of IDT is tilted from the normal of PnC by an angle $\alpha$. The scale bar is 100 μm. **c**, Topographic and **d**, TMIM-Ch1 images inside the yellow dashed box in (b). Data are taken on the sample with $R = 1.90$ μm at $f = 1.02$ GHz. The scale bar is 20 μm.

Fig. 2a shows a schematic illustration of 3D band structure of acoustic graphene, a prerequisite to emulate the Klein tunneling effect. In the following, we will follow the convention in condensed matter physics to describe the phononic system. The unit cell (dashed rhombus in Fig. 1a) of the honeycomb lattice consists of two sets of triangular sublattices, leading to the formation of linearly dispersed Dirac cones near the *K* and *K′* points of the Brillouin zone (Fig. 2b, where the blue dots correspond to the reciprocal lattice sites). For single-layer graphene with sublattice pseudospin degree of freedom, the electron wavefunction at the barrier interface perfectly matches the corresponding hole wavefunction, which is known as the charge-conjugation symmetry. Fermions in single-layer graphene thus exhibit a chirality that resembles the half-integer spinor wavefunctions in quantum electrodynamics (QED). Theoretical analysis[16,18] has shown that a massless Dirac electron normally incident on a translationally invariant potential cannot be backscattered. In other words, the electron velocity is exactly the same as in the absence of the barrier, i.e., conserved along the propagating direction. The two conditions[16,18] for Klein tunneling are thus satisfied for single-layer graphene. Because of the analogy between electronic and phononic systems, Klein tunneling is also expected for the acoustic graphene in our study.

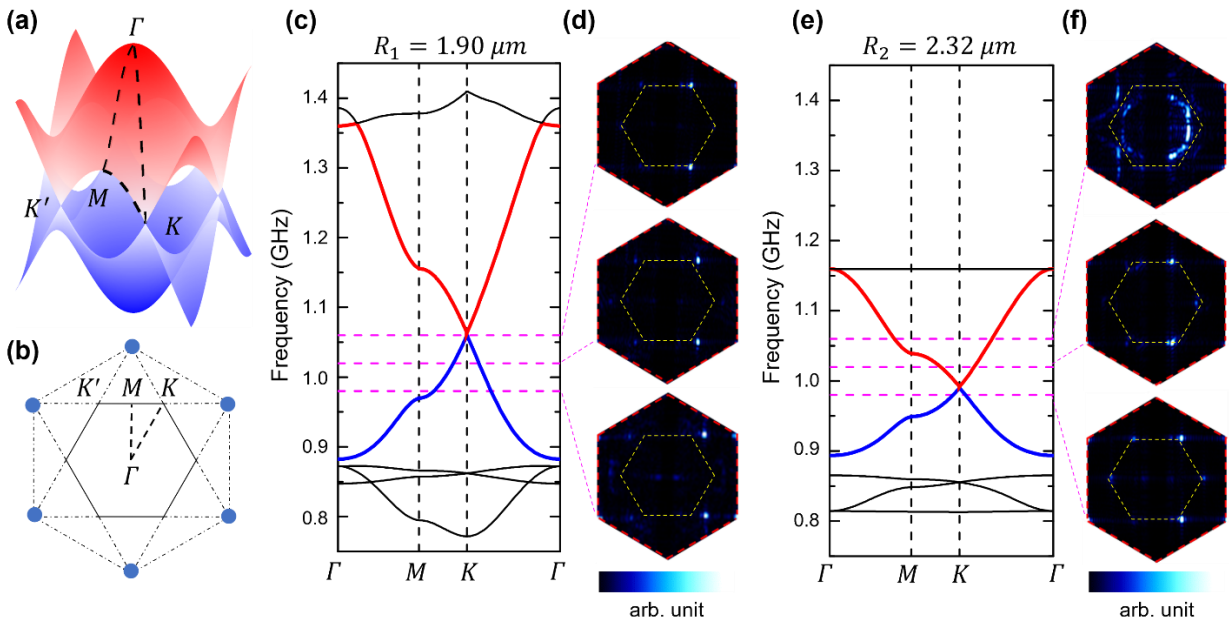

**Fig. 2| Simulated and measured phononic bands of two PnCs. a,** Schematic 3D band structure of acoustic graphene. **b,** Reciprocal-space map of the honeycomb lattice. The solid hexagon is the first Brillouin zone, with the high-symmetry points labelled in the map. The blue dots represent the reciprocal lattice sites. **c,** Simulated band structure of Sample #1 with $R_1 = 1.90$ μm. **d,** *k*-space maps of Sample #1 obtained by FFT of the TMIM data at 1.06, 1.02, and 0.98 GHz, as marked in (c). The first Brillouin zones are denoted by yellow dashed hexagons. **e** and **f,** Same as (c) and (d) for Sample #2 with $R_1 = 2.32$ μm.

A major advantage of the snowflake pattern lies in the multiple parameters for adjusting phononic bands[31], such as the length *R* and width *w* of six arms and the lattice constant $a_0$. Here we vary $R$ to tune the position of the Dirac point, while keeping $w = 1$ μm and $a_0 = 5.2$ μm unchanged in the design. Fig. 2c shows the acoustic band structure of Sample #1 with $R_1 = 1.90$ μm simulated by finite-element modeling (FEM, see Experimental Procedures), where the $K$ and lower $M$ points are situated at 1.06 and 0.97 GHz, respectively. The full band structure of the phononic crystal starting from zero frequency, as well as the mode profiles in the vicinity of the Dirac cone, are shown in Supplemental Information Section S4. Note that the Dirac-like linear dispersion no longer holds as the frequency approaches the $M$ point. For comparison, we obtain the $k$-space maps by Fourier transformation of the TMIM data. For simplicity, we will only present results near the 1st Brillouin zone and ignore FFT peaks in higher-order zones (Supplemental Information Section S3). The FFT images in Fig. 2d indicate that the Dirac point is indeed at 1.06 GHz and the peaks move away from the $K$ point at decreasing frequencies. Note that the sharp boundary of a rectangular real-space image will lead to the missing of information on the $x$-axis and $y$-axis of FFT data[33]. The third equivalent $K$ point on the $x$-axis is thus very weak due to such boundary effects. For Sample #2 with $R_2 = 2.32$ μm (Fig. 2e), the calculated $K$ and upper $M$ points are at 1.00 and 1.04 GHz, respectively. The FFT images in Fig. 2f confirm that $f = 0.98$ GHz is inside the valence band and 1.02 GHz inside the conduction band. At $f = 1.06$ GHz, the chemical potential is well above the $M$ point. Correspondingly, the FFT image displays a semicircle within the 1st Brillouin zone, in good agreement with the simulated band structure. Complete FFT data for Samples #1 and #2 can be found in Supplemental Information Section S5.

We now move on to Sample #3 formed by sandwiching a section of PnC with $R = R_1$ (Dirac point at 1.06 GHz) between two sections with $R = R_2$ (Dirac point at 1.00 GHz). For an excitation frequency between the two Dirac points, e.g., $f = 1.02$ GHz, the band alignment in Fig. 3a closely mimics the NPN configuration in gated graphene devices[19-22]. Here, the electron-like quasiparticle with energy $E = 20$ MHz in the N-type region sees an effective potential barrier of $V_0 = 60$ MHz in the P-type region. Due momentum-pseudospin locking, Dirac quasiparticles backscattered to the other sublattice under normal incidence would violate the conservation of pseudospin[16,18]. The expected transmission across the NPN heterojunction is thus unimpeded by the classically forbidden energy barrier. In our experiment, the TMIM-modulus $\left(\sqrt{V_{Ch1}^2 + V_{Ch2}^2}\right)$ image in Fig. 3b indicates that the strength of the elastic wave is comparable on both sides of the potential barrier. For quantitative analysis, we perform FFT of the TMIM data in all three sections (Supplemental

Information Section S6) and plot the FFT line profiles along the $K'$-$K$ direction in Fig. 3c. Since we use the same tip and electronics to take the TMIM image and crop the same frame for Fourier transformation, the ratio of FFT peak heights between the two $n$-type sections provides a good measure of the transmission of acoustic wave, which is near unity from Fig. 3c. Note that FFT peaks also appear near the $K'$ valley, which are associated with the reflected wave due to impedance mismatch at the sharp interface between the PnC and unpatterned AlN membrane. Theoretically, wave reflection at the PnC boundary can be suppressed by using a tapered structure[34]. In reality, however, sufficient reduction of reflected waves requires a very long taper (Supplemental Information Section S7), which is not practical in our current design. In addition, using network analysis of the transmission and reflection (Supplemental Information Section S8), one can show that multiple reflections at the PnC boundary do not affect the measurement of transmission[32]. The observation of unity transmission from N1 to N2 is also indicative of transparent hetero-interfaces without internal transmission and reflection (Supplemental Information Section S8).

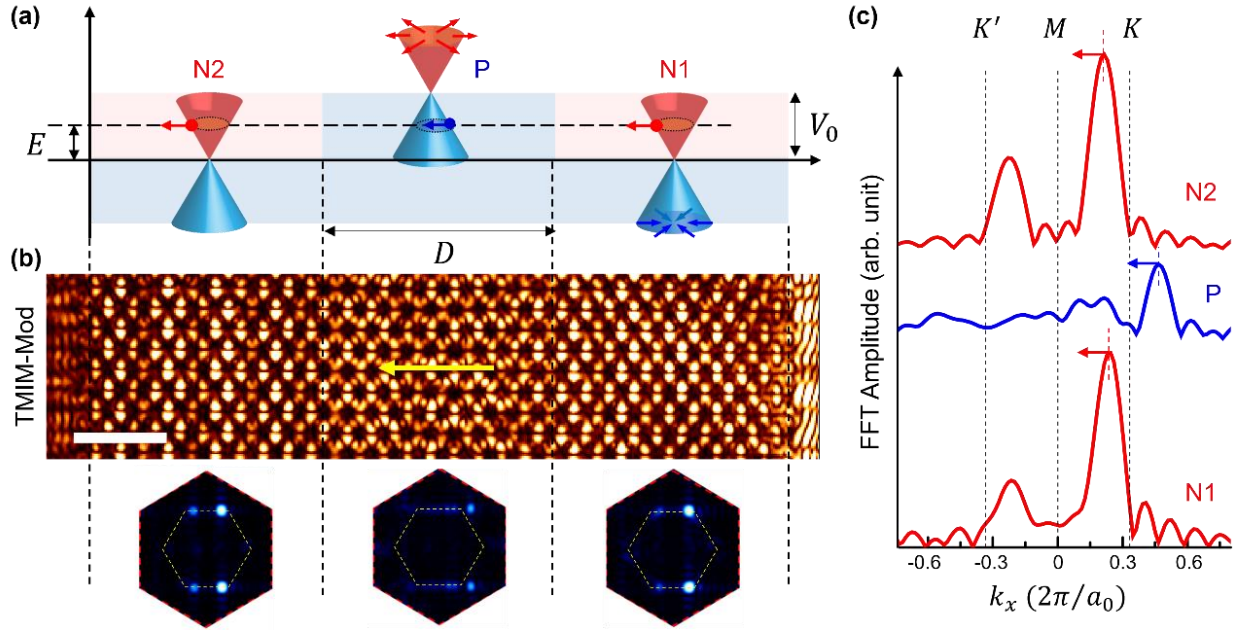

**Fig. 3| Acoustic Klein tunneling across the NPN sample. a,** Schematic of Klein tunneling of a Dirac quasiparticle with energy $E$ across a potential barrier of width $D$ and height $V_0$. The three sections of the heterostructure are labeled by N1, P, and N2. Arrows represent the direction of pseudospin. **b,** TMIM modulus image and FFT maps in each section of Sample #3 taken at $f = 1.02$ GHz. The scale bar is 20 μm. **c,** FFT line profiles along the $K'$–$K$ direction in all three sections of the NPN sample. The peak in the $P$ region corresponds to the propagating acoustic wave in the hole-like band. The peaks near the $K'$ points are associated with waves reflected from the boundary of PnC.

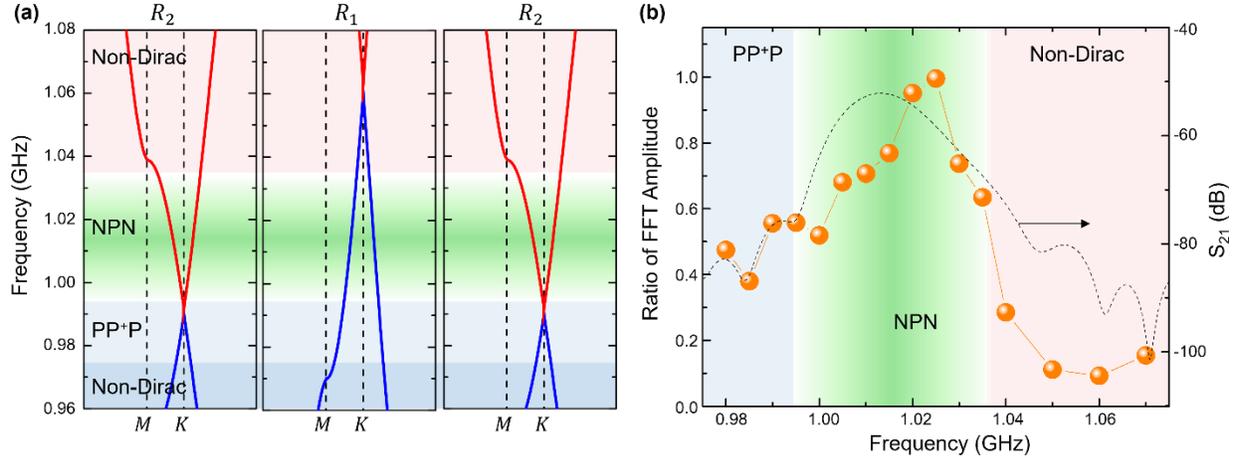

**Fig. 4| Frequency-dependent transmission under different band alignment conditions. a,** Close-up band alignment diagrams of Sample #3 near the $K$ points. **b,** Ratio of the FFT amplitude between the two $R = R_2$ sections, which is a measure of the transmission rate across the energy barrier. The transmission coefficient $S_{21}$ measured by a network analyzer is also shown for comparison.

Fig. 4a depicts the band alignment of Sample #3 near the $K$ points. Within our IDT passband of 0.98 – 1.07 GHz, there exist three regimes for the device. Below 1.00 GHz, all three sections of the sample are in the hole-like band and the configuration is denoted as PP$^+$P. In the frequency range between 1.00 and 1.03 GHz, the bands have NPN-like alignment, and Klein tunneling is expected to occur. For frequency above 1.035 GHz, the deviation from linear Dirac dispersion becomes significant in the $R = R_2$ section, i.e., the system becomes non-Dirac, and the conditions for Klein tunneling no longer exist. The raw TMIM data in the three regimes, as shown in Supplemental Section S9, are consistent with this analysis. In particular, in the Klein tunneling regime, the wavefronts in all three sections are uniform and parallel to the propagation direction. In contrast, for both the PP$^+$P and non-Dirac regimes, the wavefronts are distorted and less uniform across the sample. For quantitative evaluation of the frequency-dependent wave transmission, we plot the ratio of FFT peak amplitudes between the two $R = R_2$ sections in Fig. 4b. The ratio is around 0.5 in the PP$^+$P configuration, rises to ~ 1 in the NPN regime, and drops sharply to below 0.2 for the non-Dirac dispersion. For comparison, we also show the time-gated transmission coefficient $S_{21}$ measured by a network analyzer between the two IDTs. Apart from the slight (~ 5 MHz) difference between the frequency of $S_{21}$ peak and that of maximum FFT ratio, which may be due to extrinsic effects in the RF circuit, the image analysis from TMIM data is in good agreement with the terminal-to-terminal wave transmission. We note that near-unity transmission is only observed at the center of the NPN regime. In the snowflake design, the two sections with $R = R_1$ and $R = R_2$ exhibit very different bandwidths of the linear dispersion band, resulting in considerably different Fermi velocity (proportional to the slope *df/dk* at the *K* point) of the quasiparticle. While theoretically this does not change the unity transmission at normal incidence (see analysis below), it may, together with the multiple reflections at the PnC boundaries, affect the wave propagation close to the Dirac point. Further investigations are needed to understand the mechanism of reduction in wave transmission away from the center of the NPN regime.

Finally, we investigate the angular dependence of Klein tunneling in our phononic devices. For quasiparticle tunneling through a rectangular energy barrier, the projection of its momentum parallel to the interface is conserved since translational invariance along this direction is not affected by the interface[16,18]. The consequence of this conservation law is that the transmission coefficient is strongly dependent on the incident angle $\phi$. In other words, wave transmission is expected to be perfect only at $\phi = 0°$ and drop considerably as $\phi$ increases. To test this prediction, we fabricate 6 devices with the same heterostructure as that in Sample #3 but different launching angle $\alpha$ of the emitter IDT (Supplemental Information Section S10). The inset of Fig. 5a shows the schematic of a particular device Sample #4-3 with $\alpha = 19.5°$. By analyzing the relevant wavevectors in Fig. 5a, we can calculate the incident angle $\phi$ and refracted angle $\theta$ of group velocity on the *n*-type and *p*-type sides of the interface, respectively. Detailed *k*-space analysis of the wavevectors can be found in Supplemental Information Section S10, where all angles are tabulated for clarity.

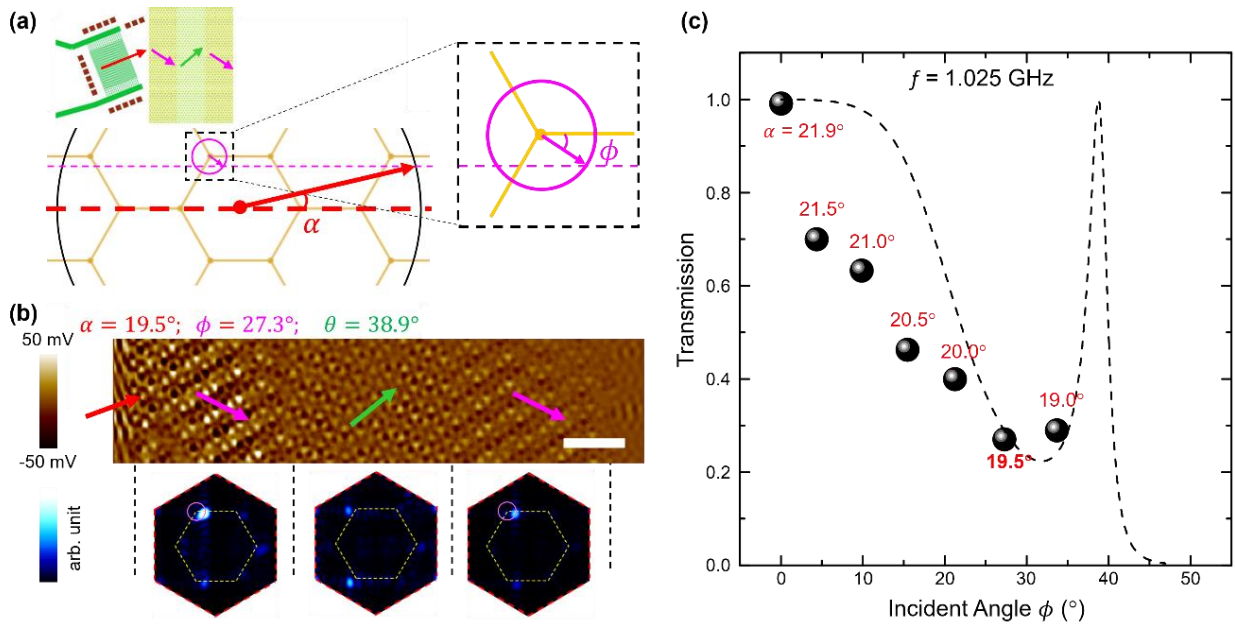

**Fig. 5| Angular dependence of acoustic Klein Tunneling. a**, Reciprocal space analysis of the incident angle $\phi$. The hexagons denote the Brillouin zones. The magenta circle represents the isotropic dispersion near the Dirac point. The inset in the upper left shows the device configuration. **b**, TMIM-Ch1 image and FFT maps of all three sections of Sample #4-3 taken at 1.025 GHz. Arrows indicate the direction of group velocity in each section. The scale bar is 20 μm. **c**, Energy transmission rate as a function of the incident angle. The IDT angles are labeled next to the data points. The dashed line is the theoretical curve. The peak at $\phi = 38.8°$ corresponds to the first-order resonant condition.

The TMIM-Ch1 image taken at $f = 1.025$ GHz and corresponding $k$-space maps of all three sections in Sample #4-3 are displayed in Fig. 5b. The theoretically calculated directions of group velocity in each section are marked in the image. A notable feature is that the refraction of Dirac quasiparticles across the heterojunction does not follow the conventional Snell's law, which is the foundation of the Veselago focusing effect[35,36]. The complete set of TMIM data for all 6 devices are included in Supplemental Section S11. Strikingly, a small change of the IDT launching angle within 3° (from 21.9° to 19.5°) leads to significant reduction in the wave transmission through the NPN heterostructure, as evident from the TMIM data. The positions and strengths of the FFT peaks also confirm the direction of wavevectors and the attenuation of transmitted waves. In Fig. 5c, we plot the energy transmission rate (square of the ratio between FFT amplitude of the peaks near $K$ points in N2 and N1) as a function of the incident angle and compare the results with theoretical analysis (Supplemental information Section S12). The substantial drop of transmission under oblique incidence is clearly observed in our experiment, which can be utilized for angle-resolved filtering applications. It should be noted that ultra-fine control of the IDT angle is needed to reach the first-order resonant condition with perfect transmission at the oblique incidence $\phi = 38.8°$ (Supplemental Information Section S12), which is very difficult for our current device design.

## Conclusions

In summary, we report the observation of Klein tunneling of GHz elastic waves in suspended piezoelectric membranes with graphene-like phononic structures. The real-space profiles of the elastic waves are visualized by TMIM, from which the reciprocal space information can be extracted by Fourier transformation and compared with simulated band structures. The near-unity transmission through an energy barrier under normal incidence is vividly observed in the NPN-like heterostructure, whereas the transmitted wave is strongly attenuated outside this regime. By controlling the launching angle of elastic waves from the emitter IDT, we also demonstrate the angular dependence of Klein tunneling in the same metamaterial design. The realization of Klein tunneling in the GHz regime has profound implications for classical and quantum acoustic devices.

For microwave signal processing, robust near-unity normal transmission is important for filtering applications. The strong angular dependence of the acoustic transmission demonstrated here will help suppress stray waves that unintentionally propagate on the device. As GHz phonons are widely utilized as information carriers in quantum computation systems, the capability of narrow-band filtering and angular waveguiding by Klein tunneling structures is desirable for the precise control of quantum transduction and information transportation. Finally, while the experimental demonstration is performed at ~ 1 GHz in this work, the device fabrication and TMIM imaging can be easily extended to the 3 ~ 6 GHz regime[37,38] for wireless telecommunication and quantum acoustic applications. In all, our work paves the way to exploit exotic high-energy-physics phenomena for classical waveguiding and integrated phononic circuit applications.

# Experimental Procedures

**Resource Availability.**

- Lead Contact: Further information and requests for resources and reagents should be directed to and will be fulfilled by the lead contact, Keji Lai (kejilai@physics.utexas.edu).
- Material Availability: This study did not generate new unique reagents.
- Data and Code Availability: The data sets generated during the current study, and/or analyzed during the current study, are available from the corresponding author upon reasonable request.

**Device fabrication.** The 800-nm-thick c-axis-oriented polycrystalline AlN films were grown by magnetron sputtering on Si wafers (from Carnegie Mellon University Claire & John Bertucci Nanotechnology Laboratory). The surface orientation is confirmed by XRD measurements. The phononic crystals consist of hexagonal arrays of etched six-armed snowflake patterns. The interdigital transducers (IDTs) were formed by the deposition and liftoff of 45-nm Al films. The suspended AlN devices were released from silicon substrates with isotropic $XeF_2$ etcher.

**Numerical simulation.** The numerical simulation is conducted by the commercial COMSOL Multiphysics software based on the finite-element method. The 'Piezoelectric Effect Multiphysics' module is applied, which couples the 'Solid Mechanics' module and 'Electrostatics' module. The Bloch boundary conditions are imposed on the boundaries of unit cells in the simulation of band structures. The material properties used in the simulations are as follows, $c_{11} = c_{22} = 375$ GPa, $c_{12} = 125$ GPa, $c_{13} = c_{23} = 120$ GPa, $c_{33} = 435$ GPa, $c_{44} = c_{55} = 118$ GPa, $c_{66} = (c_{11} - c_{12})/2 = 125$ GPa, $e_{31} = e_{32} = -0.58$ C m$^{-2}$, $e_{33} = 1.55$ C m$^{-2}$, $e_{15} = e_{24} = -0.48$ C m$^{-2}$ and $\rho = 3,100$ kg m$^{-3}$.

**Transmission-mode microwave impedance microscopy (TMIM)**. The TMIM setup is implemented on an atomic-force microscopy platform (ParkAFM, XE-70). The shielded cantilever probe (Model 5-300N) is commercially available from PrimeNano Inc. Details of the TMIM experiment can be found in Ref. 29. All measurements are performed at room temperature.

# Supplemental information description

More details on Structural characterization of the devices, Momentum-matching condition, Example TMIM and corresponding FFT images, Full band structure, Complete FFT images for Samples #1 and #2, TMIM and FFT images under Klein tunneling, Complete frequency dependent

TMIM images for Sample #3, Direction of wavevectors in Sample #4, Complete set of angular dependent TMIM images, and Theoretical analysis on the angular dependence of Klein Tunneling.

## Acknowledgements

This work was primarily supported by the National Science Foundation Division of Engineering Awards ECCS-2221822 and ECCS-2221326, and partially by the Gordon and Betty Moore Foundation, grant DOI 10.37807/gbmf12238 and the Welch Foundation grant F-1814. The device fabrication work was carried out in part at the Singh Center for Nanotechnology, part of the National Nanotechnology Coordinated Infrastructure Program, which is supported by the National Science Foundation grant NNCI-1542153.

## Author contributions

Q.Z., C.J. and K.L. conceived the project. Q.Z. and Y.J. fabricated the phononic devices and performed band-structure simulations. D.L. performed the TMIM imaging and data analysis. X.Z. and S.J. contributed to the TMIM data analysis. C.W. contributed to the phononic crystal design. D.L. and K.L. drafted the manuscript with contributions from all authors. All authors have given approval to the final version of the manuscript.

## Declaration of Interests

The authors declare no competing interests.

# Supplemental Information

# Klein Tunneling of Gigahertz Elastic Waves in Nanoelectromechanical Metamaterials

Daehun Lee[1,†], Yue Jiang[2,†], Xiaoru Zhang[1], Shahin Jahanbani[1,3], Chengyu Wen[4], Qicheng Zhang[5,6,*], A. T. Charlie Johnson[2,7,*], Keji Lai[1,*]

[1] Department of Physics, University of Texas at Austin, Austin, TX, USA
[2] Department of Physics and Astronomy, University of Pennsylvania, Philadelphia, PA, USA
[3] Present address: Department of Physics, University of California, Berkeley, CA, USA
[4] Department of Electrical and Systems Engineering, University of Pennsylvania, Philadelphia, PA, USA
[5] Research Center for Industries of the Future, Westlake University, Hangzhou, Zhejiang 310030, China
[6] Key Laboratory of 3D Micro/Nano Fabrication and Characterization of Zhejiang Province, School of Engineering, Westlake University, Hangzhou, Zhejiang 310030, China
[7] Department of Materials Science and Engineering, University of Pennsylvania, Philadelphia, PA, USA

[†] These authors contributed equally to this work

* E-mails: zhangqicheng@westlake.edu.cn; cjohnson@physics.upenn.edu; kejilai@physics.utexas.edu

## S1. Structural characterization of AlN-based phononic crystal

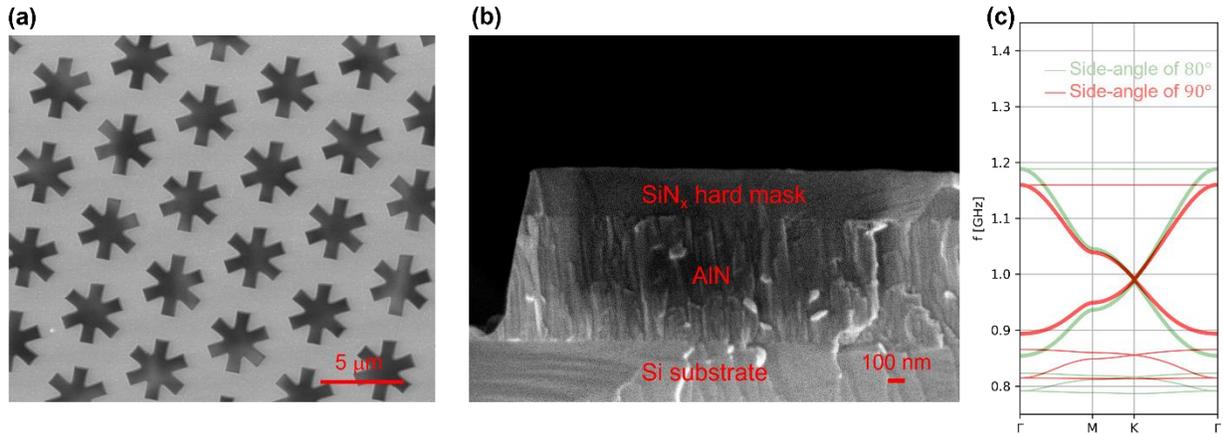

**Fig. S1 | Structural Characterization of Microfabricated Phononic Crystals. a,** Top-view SEM image of the snowflake-like pattern. **b**, Cross-sectional view of the AlN thin film before being released from the Si substrate. **c**, Simulated band structures for a phononic crystal with an ideal 90° side wall and the actual device with an 80° side wall.

The quality of the microfabricated phononic crystals in this study is examined by SEM measurements. As seen in Fig. S1a, the top-view SEM image indicates that the boundaries of the snowflake-like patterns are sufficiently sharp, confirming the high quality of our sample. Note that atomic-scale disorders will not affect the PnC band structures due to the micron-sized features. The cross-sectional SEM in Fig. S1b shows that the angle of the etched side wall is around 80°, consistent with previous reports [S1, S2]. We have investigated the effect of such imperfections on the PnC band structure by the finite element method (FEM). As shown in Fig. 1c, while this microfabrication imperfection indeed leads to a slight change to the band structure, the range of frequency with Dirac-like dispersion is almost unaffected. As a result, we will not include this minor effect in future discussions.

## S2. Momentum-matching condition

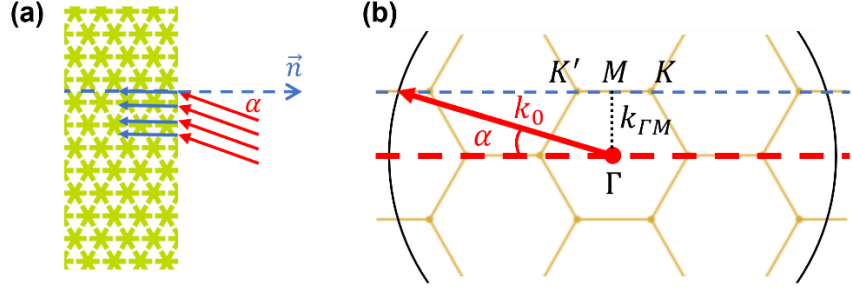

**Fig. S2 | Momentum-matching condition. a,** Illustration of the phononic crystal (PnC) with snowflake patterns, where $\vec{n}$ is the direction normal to the interface between PnC and unpatterned region. **b,** Angular selection rule in the reciprocal space. The hexagons represent the extended Brillouin zones. The blue dashed line is parallel to the PnC normal. The black circle denotes the isotropic dispersion of the polycrystalline AlN membrane. $\alpha$ represents the launching angle from the interdigital transducer (IDT) that satisfies the momentum-matching condition at $K$ or $K'$ point.

To ensure an efficient transmission of the elastic wave from the pattern-free membrane to the phononic crystal, the momentum parallel to the PnC boundary must be conserved. In this work, the desired wavevector is near the $K$ or $K'$ point of the Brillouin zone such that the wave can transmit normally into the crystal. As a result, the momentum conservation requires that $k_{\Gamma M} = k_0 \sin \alpha$, where $k_{\Gamma M}$ is the wavevector between $\Gamma$ and M points and $k_0$ is the wavevector for the pattern-free AlN membrane determined by the IDT aperture. As illustrated in Fig. S2, this condition is met when the blue dashed line intersects with the black circle. Using the PnC design parameters, one can show that the angular selection rule is satisfied at $\alpha = 21.9°$, as labeled in Fig. S2b. It is worth noting that, while $\alpha = 0°$ is also a possible solution, the small mode profile overlap weakens the overall transmission and thus is not preferred.

## S3. Example TMIM and corresponding FFT images.

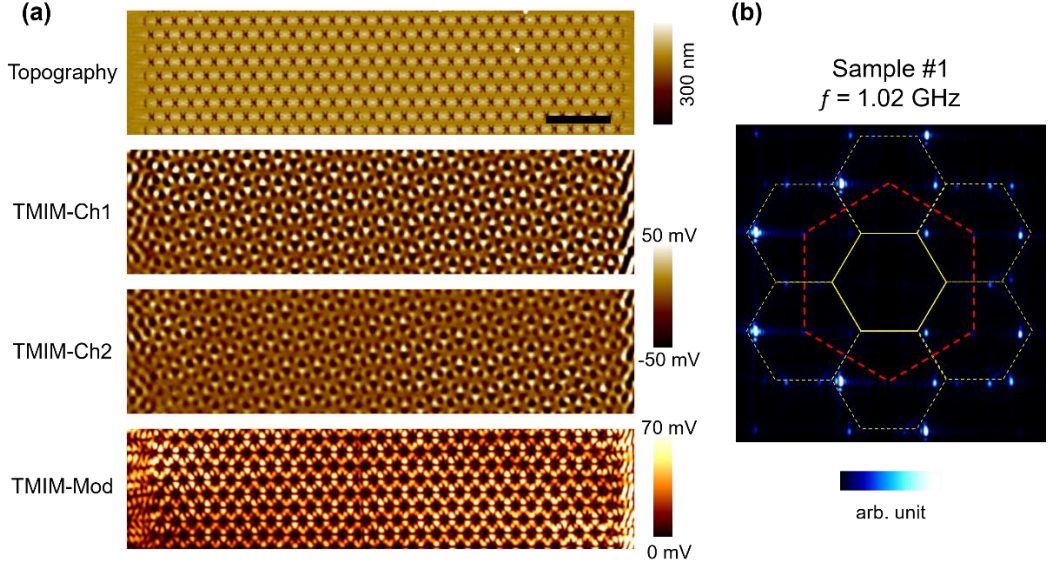

**Fig. S3 | Example TMIM and FFT images. a,** Top to bottom: Simultaneously taken topography, TMIM-Ch1/Ch2 images, as well as the calculated TMIM modulus image for Sample #1 at $f$ = 1.02 GHz. The scale bar is 20 μm. **b**, FFT image of $V_{Ch1} + i * V_{Ch2}$. The solid yellow hexagon is the 1st Brillouin zone, and the dashed hexagons are the higher-order zones. The red dashed hexagon is defined by the reciprocal lattice sites of the snowflakes, which is presented in the main text.

Fig. S3a shows examples of the experimental data presented in this work, including the surface topography, TMIM-Ch1 and -Ch2 signals, and the TMIM modulus ($\sqrt{V_{Ch1}^2 + V_{Ch2}^2}$) images for Sample #1 taken at $f$ = 1.02 GHz. As stated in the main text, the $k$-space information can be extracted through Fast-Fourier Transformation (FFT) of the combined TMIM signals $V_{Ch1} + i * V_{Ch2}$, which is shown in Fig. S3b for the same data in Fig. S3a. Here, the first Brillouin zone (BZ, solid yellow hexagon) is defined by the reciprocal lattice sites of the snowflake pattern. As seen from the raw data in Fig. 1d, there exist regions with alternating positive (bright) and negative (dark) TMIM signals within each unit cell. This is because the TMIM is sensitive to the phase of displacement fields. These fine features within the snowflake unit cell may introduce additional FFT peaks in various BZs. As a result, it is possible that the FFT spots are stronger at the $K$ points of higher order zones than that of the first zone under certain conditions. We would like to emphasize that the desired information from the FFT analysis is the overall strength of the acoustic waves across the phononic crystals. Therefore, it is sufficient to confine our analysis within the 1st BZ and compare the relative FFT peak values here.

## S4. Full phononic band structure and simulated mode profiles

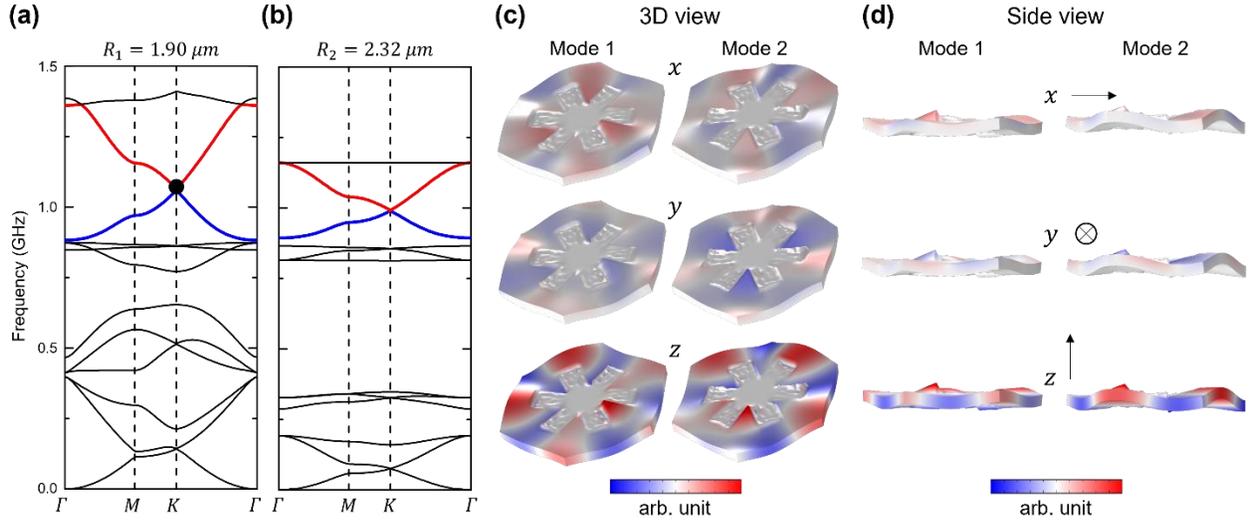

**Fig. S4 | Full band structure of the phononic crystal. a,** Full phononic band structures of Sample #1 with $R_1 = 1.90$ μm and **b**, Sample #2 with $R_1 = 2.32$ μm in the main text. **c**, Simulated surface electric potential on the left and right sides of and exactly at the Dirac point, as labeled in Panel a. The inversion of mode profiles when crossing the Dirac cone is evident in the unit-cell simulation. The two degenerate modes at the $K$ point (1.06 GHz) are linear combinations of the pseudospin eigenmodes and opposite in phase.

The complete band structure starting from zero frequency for Samples 1 and 2 in the main text are shown in Figs. S4a and S4b. In Fig. S4c, we show the simulated surface electric potential, which is mostly proportional to the out-of-plane $z$-displacement, of the modes on the left and right sides of the Dirac point. It is clear that the patterns of the two pseudospin modes are inverted when crossing the Dirac cone. The unit-cell simulation provides a vivid demonstration that the electron-like wavefunction above the Dirac cone matches the hole-like wavefunction below the Dirac cone with opposite momenta (with respect to the Dirac point), which is critical to the realization of Klein tunneling. Finally, we also include the two degenerate modes exactly at the $K$ point, which are linear combinations of the pseudospin modes and opposite in phase.

## S5. Complete FFT images for Samples #1 and #2.

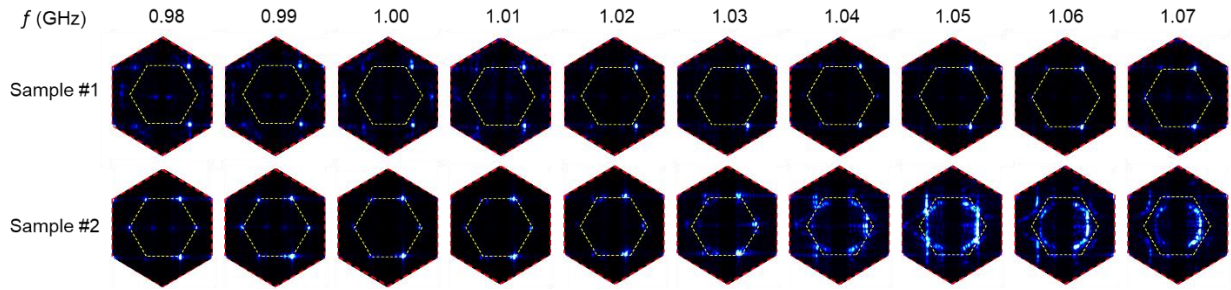

**Fig. S5 | Complete FFT images for Samples #1 and #2.** Images are shown for the IDT passband of 0.98 – 1.07 GHz, beyond which the elastic wave generation is not efficient. The yellow dashed hexagons denote the 1$^{st}$ Brillouin zones.

The complete $k$-space maps for Samples #1 and #2 are shown in Fig. S5. For Sample #1, the FFT peaks are moving towards the $K$ point as the frequency increases from 0.98 to ~ 1.06 GHz, i.e., the chemical potential is in the hole-like band. For Sample #2, as the frequency increases, the FFT peaks near the $K$ point gradually move toward the $M$ point from 0.98 to 1.03 GHz. After $f =$ 1.04 GHz, the spots become dispersive and turn into a semicircle within the 1$^{st}$ Brillouin zone, consistent with the simulated band structure in Fig. 2e in the main text.

## S6. TMIM and FFT images under Klein tunneling.

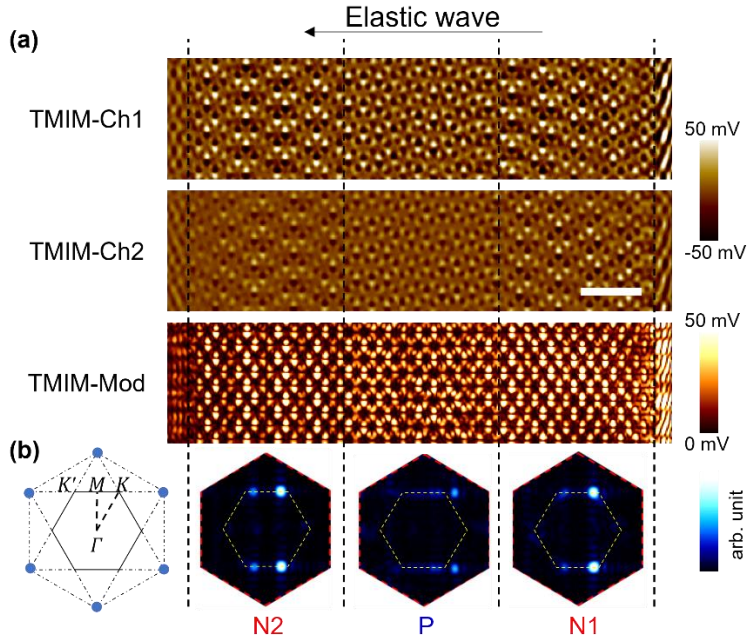

**Fig. S6 | TMIM and FFT images for Sample #3 under Klein Tunneling. a,** Top to bottom: TMIM-Ch1, -Ch2, -Mod images of the NPN Sample #3 taken at $f = 1.02$ GHz. The scale bar is 20 μm. **b,** Schematic (left) and FFT images (right) of the $k$-space maps in all three sections of the sample.

The complete TMIM images in Fig. 3 of the main text are shown in Fig. S6a. The $k$-space maps in all three sections of the NPN Sample #3 taken at $f = 1.02$ GHz are displayed in Fig. S6b. Klein tunneling across the energy barrier is clearly observed in both real-space and reciprocal-space images.

## S7. Comparison between different PnC boundaries

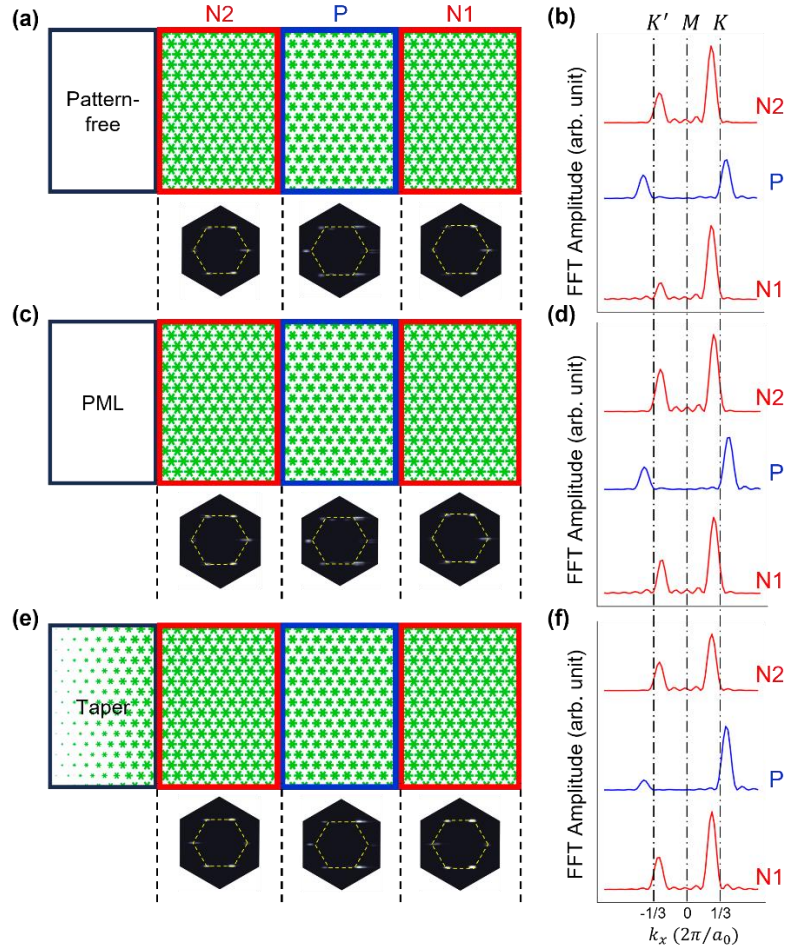

**Fig. S7 | Boundary reflection of three different conditions. a,** Schematic of NPN heterostructure terminated by pattern-free AlN membrane and *k*-space maps of corresponding regions with an incident wave from the left. **b,** FFT line profiles along the $K'$–$K$ direction in all three sections. **c** and **d**, same as a and b for PML-terminated sample. **e** and **f**, same as a and b for a tapered structure that terminates the sample. Note that wave reflection (peak near the $K'$ valley) is seen in all three cases.

   We conducted a comprehensive analysis of the wave reflection at the boundary of the NPN heterostructures: (1) pattern-free AlN membrane in Fig. S7a, (2) a perfectly matching layer (PML) in Fig. S7c, (3) a tapered PnC region [S4] in Fig. S7e. It is clear from the FFT line profiles that the strength of reflected waves is not negligible in all three cases. For (2), the transmitted wave is passively absorbed by the matching layer, whereas the reflected wave is still significant since the PML does not match the acoustic impedance. For (3), the suppression of reflection is insufficient even though the length of the taper is already comparable to that of the N2 layer. Further reduction of the reflected wave may require an adiabatically long tapered structure, which is not practical in the actual experimental design.

## S8. Network analysis of transmission and reflection.

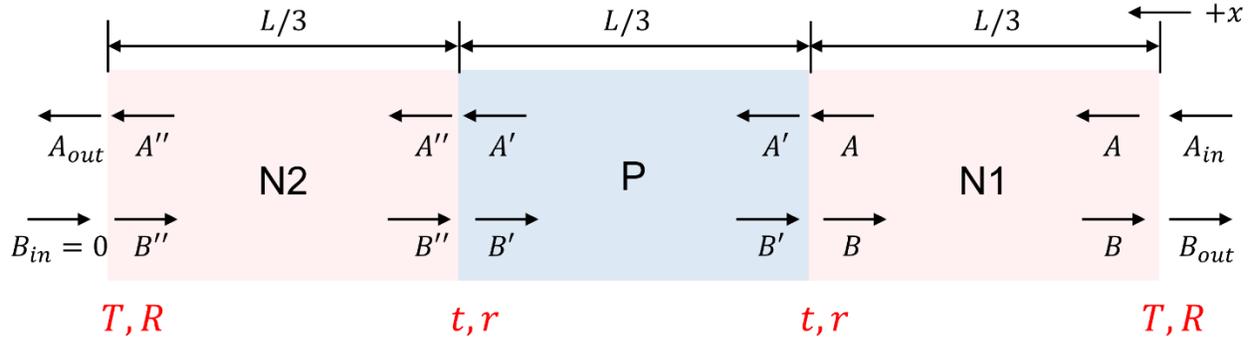

**Fig. S8 | Scattering model of the NPN heterostructure in the presence of internal and boundary reflections.** Schematic of NPN heterostructure with multiple reflections and transmissions. The same transmission ($T$) and reflection ($R$) coefficients for both PnC-membrane boundaries, as well as the same internal transmission ($t$) and reflection ($r$) coefficients for both hetero-interfaces, are assumed due to time-reversal symmetry.

Fig. S8 shows the network scattering model of the NPN heterostructure in the presence of internal and boundary reflections. In the case of transparent N-P-N hetero-interfaces ($t = 1$ and $r = 0$) for Klein tunneling, it is obvious that $A = A'' = TA_{in} + TR^2 A_{in} + TR^4 A_{in} \ldots = TA_{in}/(1 - R^2)$, which is the same as the result in our previous paper [S5] if we ignore the propagation loss. In other words, the presence of boundary reflections ($T \neq 1$ and $R \neq 0$) simply reduces the ratio between transmitted wave amplitude and incident wave amplitude by a common factor. On the other hand, if the hetero-interfaces are not transparent, the transmitted amplitude becomes

$$A/A_{in} = T[(1 + Rr + R^2 r^2 + \cdots) + t^2(Rr + Rr^3 + 2R^2 r^2 + \cdots) + t^4(R^2 + 3R^3 r + 3R^2 r^2 + \cdots)]$$

$$A''/A_{in} = Tt^2[(1 + r^2 + \cdots) + t^2(2Rr + 5R^2 r^2 + \cdots) + t^4(4Rr^3 + \cdots) + t^6(R^2 + \cdots)]$$

In general, we will not have $A = A''$ when internal reflection ($t \neq 1$ and $r \neq 0$) occurs, which contradicts with our data in Fig. 3c. In summary, the presence of boundary reflections ($T \neq 1$ and $R \neq 0$) does not affect the transmission measurement, and the presence of internal reflections ($t \neq 1$ and $r \neq 0$) is not supported by our experimental result.

## S9. Complete frequency dependent TMIM images for Sample #3.

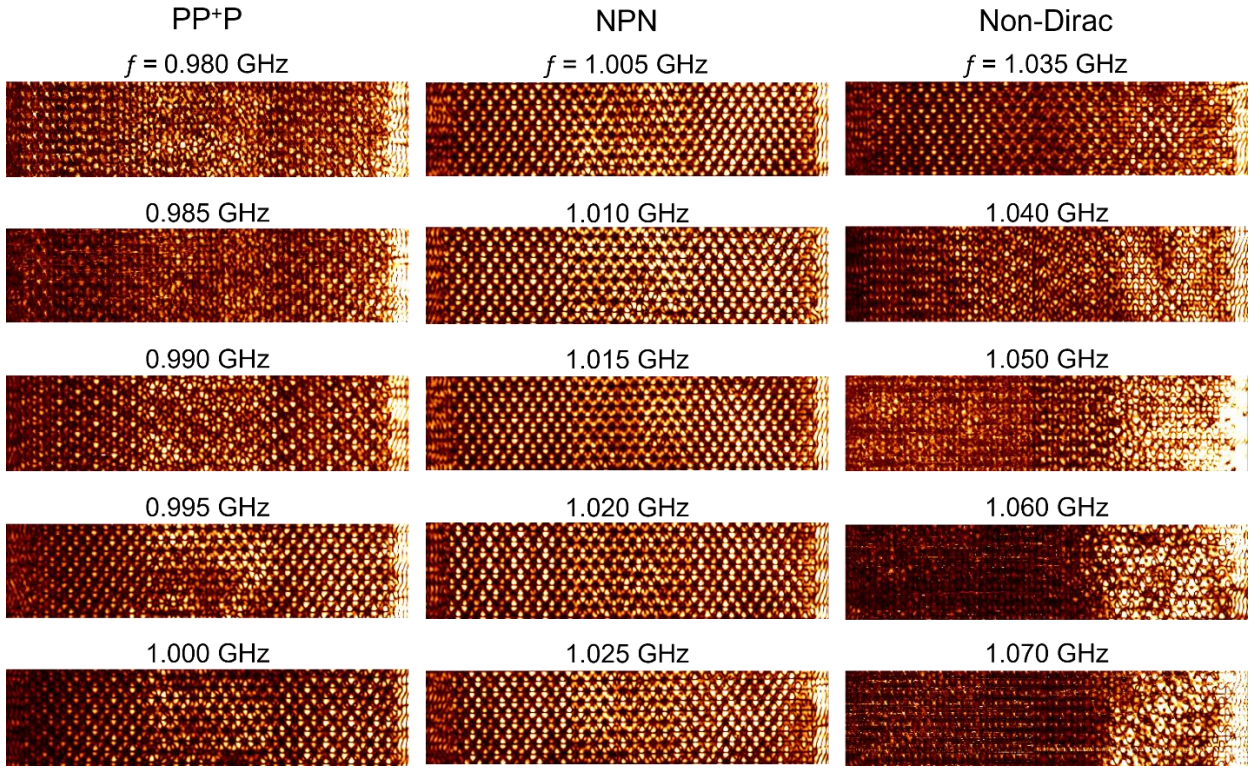

**Fig. S9 | Complete set of frequency dependent TMIM modulus images for Sample #3.** All images are 170 μm × 40 μm. The full color scales are from 0 mV to 50 mV.

The complete set of frequency dependent TMIM modulus images for Sample #3 are shown in Fig. S9. In the nominally Klein tunneling regime from 1.00 to 1.03 GHz, the wavefronts in all three sections are uniform and parallel to the propagation direction. In contrast, for both the PP$^+$P and non-Dirac regimes, the wavefronts are distorted and less uniform across the sample.

## S10. Direction of wavevectors in Sample #4.

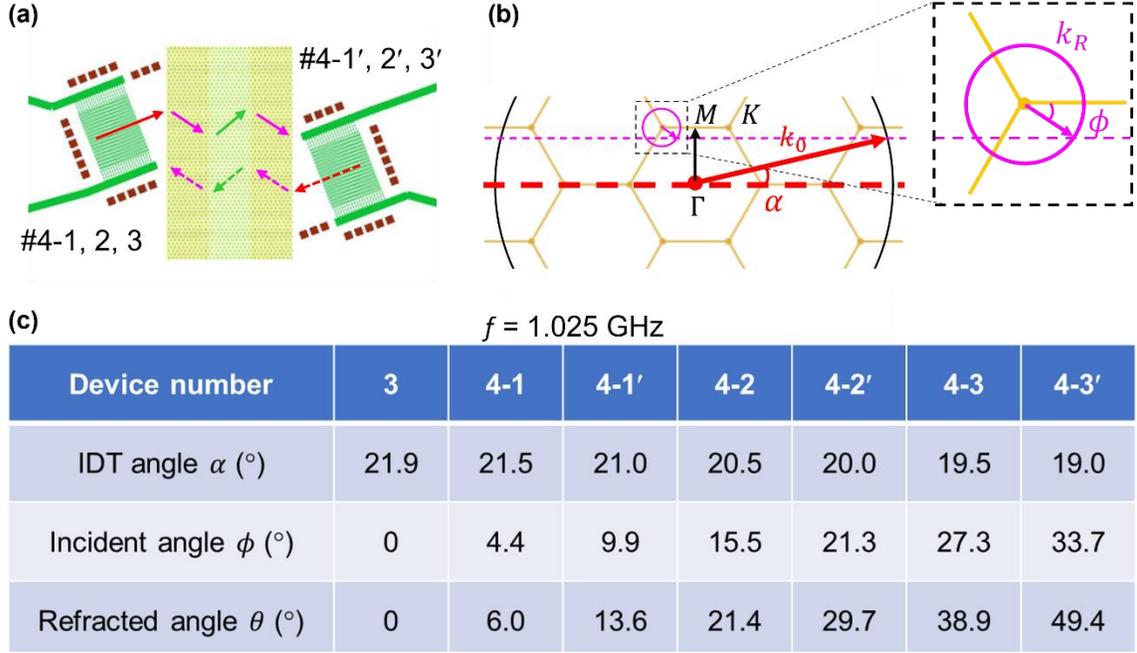

**Fig. S10 | Direction of wavevectors in Sample #4 for the angular dependence study. a,** Illustration of device structures for the angular dependence study. Samples #4-1 and #4-1′ (same for #4-2 and #4-2′, #4-3 and #4-3′) share the same PnC but differ slightly in the IDC launching angle $\alpha$. Arrows denote the direction of group velocity in each section of the sample. **b,** Schematic for calculating the incident angle $\phi$ (see text below). **c,** Table summarizing the launching angle $\alpha$, incident angle $\phi$, and refracted angle $\theta$ in all devices.

In this work, we design 6 devices (3 pairs as illustrated in Fig. S10a) for the angular dependence study. The relevant angles (all measured with respect to the *x*-axis or horizontal direction) are defined as follows – $\alpha$: IDT launching angle of elastic wave in the pattern-free membrane; $\phi$: incident angle in the *n*-type section toward the energy barrier; $\theta$: refracted angle in the *p*-type section. Note that wave refraction in our case does not follow the usual Snell's law.

Fig. S10b shows the *k*-space schematic for the PnC. Here $k_0$ is the wavevector of elastic waves generated by IDT, which is determined by the IDT aperture. $k_R$ is the frequency-dependent wavevector radius of the energy iso-surface circle centered at the *K* point. $k_{\Gamma M} = 2\pi/\sqrt{3}a_0$ is the wavevector from Γ to M. The geometric relation in Fig. S8b indicates that $k_{\Gamma M} = k_0 \sin \alpha + k_R \sin \phi$, from which the incident angle $\phi$ can be calculated at a certain IDT angle $\alpha$. The angles in Samples #3 and #4 at *f* = 1.025 GHz are summarized in the table in Fig. S10c. The calculation of refracted angle $\theta$ will be discussed in Supplemental Section 12.

## S11. Complete set of angular dependent TMIM images.

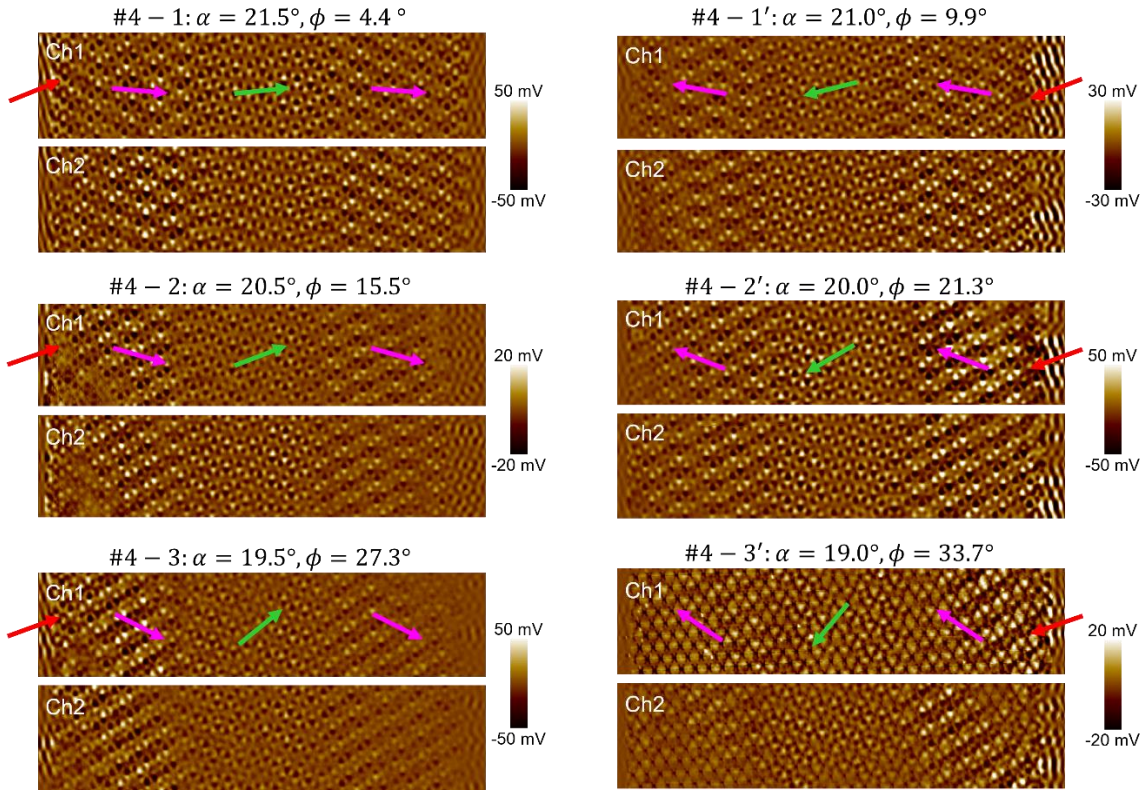

**Fig. S11 | Complete set of angular dependent TMIM images in all 6 devices taken at *f* = 1.025 GHz.** All images are 40 μm × 170 μm.

Fig. S11 shows the TMIM images in all six devices (Samples #4-1, -1′, -2, -2′, -3, -3′) in this study taken at *f* = 1.025 GHz. Arrows labeled in the TMIM-Ch1 images correspond to the direction of group velocity in each section. Angles are taken from the table in Fig. S10c. Note that the group velocity points to the same (opposite) direction as the wavevector in the *n*-type (*p*-type) PnC.

**S12. Theoretical analysis on the angular dependence of Klein Tunneling.**

In our phononic metamaterials, the interface between different sections of PnC is sharp compared to the Fermi wavelength. As a result, we can follow Ref. [S3] to calculate the transmission of a quasiparticle with energy $E$ through a rectangular potential barrier that is infinite along the y-axis,

$$V(x) = \begin{cases} V_0, & 0 < x < D, \\ 0, & \text{otherwise.} \end{cases}$$

, where $D = 10a_0$ is the width of the barrier and $V_0$ is the height of the barrier determined by the energy difference between Dirac points of p-type ($R = R_1$) and n-type ($R = R_2$) PnCs. We emphasize that the difference between Fermi velocity in the n-type ($v_{Fn}$) and p-type ($v_{Fp}$) PnCs is not negligible in our sample due to the different bandwidths of the Dirac-like cones (see Figs. 2c and 2e in the main text). The theoretical analysis in our work is thus different from that in Ref. [S3] as follows,

- Fermi wavevector in the n-type PnC: $k_F = E/\hbar v_{Fn}$, $k_x = k_F \cos\phi$, $k_y = k_F \sin\phi$
- Fermi wavevector in the p-type PnC: $q_F = (V_0 - E)/\hbar v_{Fp}$, $q_x = q_F \cos\theta$, $q_y = q_F \sin\theta$
- Conservation of momentum parallel to the interface: $k_y = q_y$
- Refracted angle: $\theta = \tan^{-1}(k_y/q_x)$, where $q_x = \sqrt{q_F^2 - q_y^2} = \sqrt{q_F^2 - k_y^2}$
- Energy transmission across the barrier, which is used for the theoretical curve in Fig. 5c in the main text:

$$T = \frac{\cos^2\phi \cos^2\theta}{\cos^2\phi \cos^2\theta \cos^2(q_x D) + \sin^2(q_x D)[1 + \sin\theta \sin\phi]^2}$$

The main difference between our analysis and that in Ref. [S3] lies in the explicit inclusion of different $v_{Fn}$ and $v_{Fp}$, which are extracted from our FEM result. As seen from the expression above, the transmission is always 1 with $\phi = \theta = 0$. In other words, the considerably different $v_{Fn}$ and $v_{Fp}$ does not affect unity transmission upon normal incidence. For oblique incidence, upon resonant conditions at $q_x D = N\pi$, $N = \pm 1, \pm 2 ...$, the barrier becomes transparent with $T = 1$. Specifically, the peak at the oblique incidence angle $\phi = 38.8°$ in Fig. 5c of the main text corresponds to the first-order resonance or $N = 1$. Note that ultra-fine control of the IDT angle is needed to reach this condition, which is difficult to meet in our design.